\documentclass[lettersize,journal]{IEEEtran}
\usepackage{amsmath, amsfonts, amssymb}
\usepackage{algorithmic}
\usepackage{algorithm}
\usepackage{array}
\usepackage{pifont}
\usepackage[caption=false,font=normalsize,labelfont=scriptsize,textfont=scriptsize]{subfig}
\usepackage{textcomp}
\usepackage{stfloats}
\usepackage{url}
\usepackage{verbatim}
\usepackage{graphicx}
\usepackage{cite}

\usepackage[bottom = 1in,top = 0.7in,left=0.6in,right=0.6in]{geometry}
\begin{document}

\title{Downlink Multiuser Communications Relying on Flexible Intelligent Metasurfaces}
\author{\small \IEEEauthorblockN{Jiancheng An\IEEEauthorrefmark{1}, Chau Yuen\IEEEauthorrefmark{1}, Marco Di Renzo\IEEEauthorrefmark{2}, M\'erouane Debbah\IEEEauthorrefmark{3}, H. Vincent Poor\IEEEauthorrefmark{4}, and Lajos Hanzo\IEEEauthorrefmark{5}}\\
\IEEEauthorblockA{\IEEEauthorrefmark{1}School of Electrical and Electronics Engineering, Nanyang Technological University, Singapore 639798\\
\IEEEauthorrefmark{2}CentraleSup\'elec, Laboratoire des Signaux et Syst\`emes, Universit\'e Paris-Saclay, 91192 Gif-sur-Yvette, France\\
\IEEEauthorrefmark{3}Center for 6G, Khalifa University of Science and Technology, P O Box 127788, Abu Dhabi, United Arab Emirates\\
\IEEEauthorrefmark{4}Department of Electrical and Computer Engineering, Princeton University, Princeton, NJ 08544 USA\\
\IEEEauthorrefmark{5}School of Electronics and Computer Science, University of Southampton, SO17 1BJ Southampton, U.K.\\
E-mail: \{jiancheng.an, chau.yuen\}@ntu.edu.sg, marco.di-renzo@universite-paris-saclay.fr, merouane.debbah@ku.ac.ae, poor@princeton.edu, lh@ecs.soton.ac.uk
}\vspace{-0.1cm}}
%\markboth{GLOBECOM}{GLOBECOM}
\maketitle
\begin{abstract}
A flexible intelligent metasurface (FIM) is composed of an array of low-cost radiating elements, each of which can independently radiate electromagnetic signals and flexibly adjust its position through a 3D surface-morphing process. In our system, an FIM is deployed at a base station (BS) that transmits to multiple single-antenna users. We formulate an optimization problem for minimizing the total downlink transmit power at the BS by jointly optimizing the transmit beamforming and the FIM's surface shape, subject to an individual signal-to-interference-plus-noise ratio (SINR) constraint for each user as well as to a constraint on the maximum morphing range of the FIM. To address this problem, an efficient alternating optimization method is proposed to iteratively update the FIM's surface shape and the transmit beamformer to gradually reduce the transmit power. Finally, our simulation results show that at a given data rate the FIM reduces the transmit power by about $3$ dB compared to conventional rigid 2D arrays.
\end{abstract}


\thispagestyle{empty}
\section{Introduction}
To achieve the challenging quality of service (QoS) demands of emerging applications, metasurface technology is regarded as one of the most promising solutions \cite{TCCN_2020_Qin_20, WC_2022_An_Codebook, JSAC_2020_Tang_MIMO}. In general, a metasurface is an artificially engineered planar surface constructed of an array of tiny meta-atoms \cite{Proc_2022_Alexandropoulos_Pervasive}. Each meta-atom is capable of independently radiating or scattering wireless signals by beneficially modifying the properties of electromagnetic (EM) waves, such as their amplitudes, phases, and polarizations \cite{WC_2022_An_Codebook}. The deployment of near-passive reconfigurable intelligent surfaces (RISs) in wireless networks has shown great potential for shaping their wireless propagation environment \cite{TCOM_2024_Yu_Environment, TCOM_2022_An_Low}. Since RISs interact with the incoming signals without requiring power amplifiers and complex signal processing, they avoid the costly radio frequency hardware of conventional active transceivers \cite{WC_2022_An_Codebook}.

Additionally, a complementary application to near-passive RISs is to harness active metasurfaces as transceivers, which are known as large intelligent surfaces (LISs) \cite{arXiv_2023_An_Stacked_mag, TSP_2018_Hu_Beyond, arXiv_2024_Liu_Stacked}. Given the large aperture operating at high frequencies, the near-field Fresnel region of an LIS can extend to hundreds of meters \cite{CM_2023_Cui_Near}. The intrinsic spherical wavefront characteristics allow it to harvest the full spatial multiplexing gain even under strong line-of-sight (LoS) propagation conditions \cite{arXiv_2023_An_Toward}. Furthermore, metasurfaces enable EM operations at an unprecedented level of resolution \cite{CM_2021_Dardari_Holographic, CM_2022_Zhang_Intelligent}, underpinning the concept of holographic multiple-input and multiple-output (MIMO) communications by modeling an LIS as a near-continuous array of a massive number of infinitesimally small antennas \cite{TWC_2022_Pizzo_Fourier, CL_2023_An_A1}.

Nevertheless, existing studies on metasurface applications in wireless communications have generally adopted rigid metamaterials \cite{JSAC_2020_Tang_MIMO, Proc_2022_Alexandropoulos_Pervasive}. Thanks to recent developments in micro/nano-fabrication and the discovery of flexible metamaterials, it has become possible to create flexible intelligent metasurfaces (FIMs) by depositing dielectric inclusions on conformal flexible substrates \cite{Sci_2015_Ni_An, SR_2016_Cheng_All}. In contrast to rigid metasurfaces, FIMs show significant potential, particularly for applications involving wave manipulation on conformal and curved surfaces \cite{NC_2016_Kamali_Decoupling}. To support dynamic and programmable 3D surface-shape morphing, the authors of \cite{NC_2022_Ni_Soft} exploited a soft microfluidic liquid metal network embedded in an elastomer controlled by an EM actuation. More recently, an FIM was built from a matrix of tiny metallic filaments \cite{Nature_2022_Bai_A}, driven by reprogrammable distributed Lorentz forces emanating from electrical currents passing through a static magnetic field\footnote{Please refer to \url{https://www.eurekalert.org/multimedia/950133} to watch a video showing the real-time morphing capability of an FIM.}. This endows the FIM with precise and rapid dynamic morphing capabilities for promptly changing its structure.

Even though FIMs show promise in many applications, their potential use in wireless networks has hitherto remained largely unexplored. However, FIMs are actually well-suited for wireless communications. Specifically, wireless channels typically experience fading effects due to multipath propagation. Multiple copies of the transmitted signal traveling along different paths with different signal attenuations, delays, and phase shifts may result in either constructive or destructive interference at the receiver \cite{BOOK_2005_Tse_Fundamentals}. Severe destructive interference can significantly degrade the channel quality and cause communication outages. Fortunately, strategically morphing the surface shape of the FIM may guarantee that the multiple copies of the signal add constructively at the FIM to boost the received signal power. This is particularly beneficial for wireless networks operating at the mmWave and THz frequencies \cite{TWC_2022_Wong_Fluid}, since the coherence distance is very small. Therefore, the FIM is expected to be a promising technology for further improving both the spectral and energy efficiency of next-generation wireless networks.

Against this background, we investigate a multiuser multiple-input single-output (MISO) communication system in which an active FIM is deployed at the base station (BS). Specifically, we formulate an optimization problem for minimizing the total transmit power at the BS by jointly optimizing the transmit beamforming vectors and the morphed surface shape of the FIM, subject to a specific signal-to-interference-plus-noise ratio (SINR) target for each user and the morphing range of the FIM. Since this problem is challenging to solve optimally due to the non-convex SINR constraints, we propose an efficient alternating optimization algorithm for iteratively updating both the surface shape of the FIM and the transmit beamforming vectors. In each iteration, the optimal transmit beamforming weights are derived in a closed form given the surface shape of the FIM, while the surface shape of the FIM is updated for increasing the SINR margin and for implicitly reducing the transmit power. Numerical results demonstrate that harnessing an FIM can substantially reduce the transmit power required at the BS to fulfill the SINR targets of the users.

\thispagestyle{empty}
\section{System Model}\label{sec2}
\begin{figure}[!t]
\centering
\includegraphics[width = 8 cm]{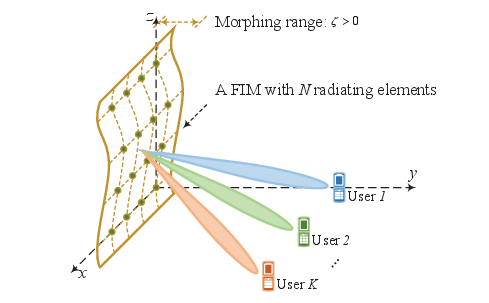}
\caption{Schematic of a multiuser MISO system, where an FIM is deployed at the BS.}\vspace{-0.1cm}
\label{fig_2}
\end{figure}

As shown in Fig. \ref{fig_2}, we consider the downlink of a multiuser MISO communication system, where a BS equipped with an FIM communicates simultaneously with $K$ single-antenna mobile users. The unstretched FIM is modeled as a uniform planar array (UPA) located on the $x$-$z$ plane. Let $N=N_{x}N_{z}$ represent the total number of transmit antennas, with $N_{x}$ and $N_{z}$ referring to the number of antenna elements along the $x$-axis and $z$-axis, respectively. We define the sets of antennas and users as $\mathcal{N} \triangleq \left \{ 1, 2, \cdots, N \right \}$ and $\mathcal{K} \triangleq \left \{ 1, 2, \cdots, K \right \}$, respectively.

In contrast to conventional communication systems relying on rigid antenna arrays, each radiating element of the FIM can be flexibly positioned along the direction perpendicular to the surface, i.e., the $y$-axis, with the aid of a controller. Specifically, let $\boldsymbol{p}_{n}=\left [ x_{n},y_{n},z_{n} \right ]^{T}\in \mathbb{R}^{3},\ \forall n \in \mathcal{N}$ represent the location of the $n$-th radiating element. Taking the first element as a reference point, we have $x_{n} = d_{x}\times \textrm{mod}\left ( n-1,N_{x} \right )$ and $z_{n} = d_{z}\times \left \lfloor \left ( n-1 \right )/N_{x} \right \rfloor$, where $d_{x}$ and $d_{z}$ denote the spacing between the adjacent antenna elements in the $x$- and $z$-directions, respectively. Furthermore, the $y$-coordinate of each radiating element can be adjusted within the maximum range allowed by the reversible deformation of the FIM \cite{Nature_2022_Bai_A}, satisfying $y_{\min} \leq y_{n}\leq y_{\max}$, $\forall n \in \mathcal{N}$, where $y_{\min}$ and $y_{\max}$ represent the minimum and maximum $y$-coordinates of each element, and $\zeta =y_{\max}-y_{\min}> 0$ is termed as the \emph{morphing range} characterizing the range of reversible deformation. We set $y_{\min} = 0$ throughout this paper. Therefore, the surface shape of the FIM is characterized by
\begin{align}
    \boldsymbol{y}=\left [ y_{1},y_{2},\cdots ,y_{N} \right ]^{T}\in \mathbb{C}^{N}.
\end{align}

Moreover, we assume that all channels experience quasi-static flat fading. Let $\boldsymbol{h}_{k}^{H}\in \mathbb{C}^{1\times N},\ \forall k\in \mathcal{K}$ denote the baseband equivalent channel spanning from the FIM at the BS to the $k$-th user. We adopt the multipath channel representation of \cite{JSTSP_2016_Heath_An} to characterize the wireless channels. Specifically, the channel $\boldsymbol{h}_{k}^{H}$ is modeled as a composite response of multiple propagation paths. For a scatterer in the far field of the FIM, the phase profile observed by the antenna array depends on the scattering elevation angle $\theta\in \left [ 0,\pi \right )$, azimuth angle $\phi\in \left [ 0,\pi \right )$, and the FIM's surface-shape vector $\boldsymbol{y}$. Specifically, the array steering vector $\boldsymbol{a}\left ( \boldsymbol{y}, \phi ,\theta \right )\in \mathbb{C}^{N}$ is given by
\begin{align}\label{eq5}
 \boldsymbol{a}\left ( \boldsymbol{y},\phi ,\theta \right )=\left [ 1,\cdots , e^{j\kappa \left ( x_{n}\sin\theta \cos\phi+y_{n}\sin\theta \sin\phi +z_{n}\cos\theta \right )}, \right.\notag\\
\left. \cdots ,e^{j\kappa \left ( x_{N}\sin\theta \cos\phi+y_{N}\sin\theta \sin\phi +z_{N}\cos\theta \right )} \right ]^{T},
\end{align}
where $\kappa =2\pi /\lambda $ represents the wavenumber, with $\lambda $ denoting the radio wavelength.

Furthermore, let $L$ represent the number of propagation paths between each user and the BS, while the corresponding path set is defined by $\mathcal{L} \triangleq \left \{ 1, 2, \cdots, L \right \}$. The complex gain of the $\ell$-th propagation path from the $k$-th user to the BS is represented by $\alpha_{k,\ell} \in \mathbb{C},\ \forall \ell \in \mathcal{L},\ \forall k\in \mathcal{K}$, while $\theta _{\ell}$ and $\phi _{\ell}$ are the elevation and azimuth angles of arrival (AoA), respectively, for the $\ell$-th propagation path. As a result, the narrowband channel $\boldsymbol{h}_{k}$ can be written as
\begin{align}
 \boldsymbol{h}_{k}\left ( \boldsymbol{y} \right )=\sum_{\ell=1}^{L}\alpha _{k,\ell}\boldsymbol{a}\left ( \boldsymbol{y},\phi_{\ell} ,\theta_{\ell} \right ),\quad \forall k\in \mathcal{K},
\end{align}
where $\alpha _{k, \ell},\ \forall \ell \in \mathcal{L},\ \forall k\in \mathcal{K}$ are independent and identically distributed (i.i.d.) circularly symmetric complex Gaussian random variables, satisfying $\alpha _{k, \ell}\sim \mathcal{CN}\left ( 0,\rho _{k, \ell}^{2} \right )$, where $\rho _{k, \ell}^{2}$ represents the average power of the $\ell$-th path for the $k$-th user \cite{JSTSP_2016_Heath_An}. Furthermore, let $\beta _{k}$ characterize the path loss between the $k$-th user and the BS. Thus, we have $\sum\nolimits_{\ell=1}^{L}\rho _{k, \ell}^{2}=\beta _{k},\ \forall k\in \mathcal{K}$.

Additionally, the BS employs a linear transmit precoding technique to send signals from its $N$ antennas to $K$ users \cite{SPM_2014_Bjornson_Optimal}. Let $s_{k}\in \mathbb{C},\ k\in \mathcal{K}$ denote the normalized information signal destined for user $k$, which are assumed to be independent random variables with zero mean and unit variance. Additionally, $\boldsymbol{w}_{k}\in \mathbb{C}^{N}$ represents the dedicated beamforming vector assigned to user $k$. Therefore, the complex baseband signal $\boldsymbol{u}\in \mathbb{C}^{N}$ transmitted from the BS can be expressed as
\begin{align}
\boldsymbol{u}=\sum_{k=1}^{K}\boldsymbol{w}_{k}s_{k}.
\end{align}

After propagating through the wireless channel, the complex baseband signal $r_{k}\in \mathbb{C},\ \forall k\in \mathcal{K}$ received by user $k$ is expressed as
\begin{align}
 r_{k} = \boldsymbol{h}_{k}^{H}\left ( \boldsymbol{y} \right )\boldsymbol{u}+n_{k} = \boldsymbol{h}_{k}^{H}\left ( \boldsymbol{y} \right )\sum_{{k}'=1}^{K}\boldsymbol{w}_{{k}'}s_{{k}'}+n_{k},
\end{align}
where $n_{k}\sim \mathcal{CN}\left ( 0,\sigma _{k}^{2} \right )$ represents the additive white Gaussian noise (AWGN) at user $k$'s receiver, with $\sigma _{k}^{2}$ being the corresponding noise power.

\thispagestyle{empty}
\section{Problem Formulation and Solution}\label{sec3}
\subsection{Power Minimization Problem Formulation}
We aim for minimizing the total transmit power at the BS by jointly optimizing the transmit beamforming vectors $\left \{ \boldsymbol{w}_{k} \right \}$ for multiple users and the surface shape $\boldsymbol{y}$ of the FIM, subject to a set of individual SINR requirements for the $K$ users and the morphing range. We assume that both the BS and users have the perfect knowledge of the CSI for all the channels. Specifically, the joint optimization problem is formulated as
\begin{subequations}\label{eq10}
\begin{alignat}{3}
& \left ( P_\mathcal{A} \right ) &\quad&\min_{\left \{ \boldsymbol{w}_{k} \right \},\, \boldsymbol{y}} &\quad& \sum_{k=1}^{K}\left \| \boldsymbol{w}_{k} \right \|^{2} \label{eq10a}\\
&&&\textrm{s.t.} & & \frac{\left | \boldsymbol{h}_{k}^{H}\left ( \boldsymbol{y} \right )\boldsymbol{w}_{k} \right |^{2}}{\sum\limits_{{k}'\neq k}^{K}\left | \boldsymbol{h}_{k}^{H}\left ( \boldsymbol{y} \right )\boldsymbol{w}_{{k}'} \right |^{2}+\sigma_{k} ^{2}}\geq \gamma _{k},\quad \forall k\in \mathcal{K}, \label{eq10b}\\
&&& & & 0\leq y_{n}\leq y_{\max},\quad \forall n\in \mathcal{N}, \label{eq10c}
\end{alignat}
\end{subequations}
where $\gamma _{k}>0$ represents the minimum SINR requirement for user $k$ to achieve its desired data rate. While the objective function of $\left ( P_\mathcal{A} \right )$ and the constraint in \eqref{eq10c} are convex, problem $\left ( P_\mathcal{A} \right )$ is quite challenging to solve optimally due to the non-convex constraint in \eqref{eq10b} \cite{SPM_2014_Bjornson_Optimal}. Additionally, the transmit beamforming vectors and the FIM's surface shape are highly coupled. Next, we will propose an efficient alternating optimization algorithm and derive the asymptotic beamforming vectors in the high signal-to-noise ratio (SNR) region.

\subsection{The Proposed Alternating Optimization Algorithm}\label{sec5_1}
The alternating optimization algorithm involves solving a pair of subproblems. Specifically, the transmit beamforming vectors and the FIM's surface shape are optimized alternately, until convergence is reached.
\subsubsection{Transmit Beamforming Optimization $\left \{ \boldsymbol{w}_{k} \right \}$ Given the FIM's Surface Shape $\hat{\boldsymbol{y}}$}
For a tentative surface-shape vector $\hat{\boldsymbol{y}}$, the wireless channel from user $k$ to the BS is determined by $\boldsymbol{h}_{k}\left ( \hat{\boldsymbol{y}} \right )=\sum\nolimits_{\ell=1}^{L} \alpha _{k,\ell}\boldsymbol{a}\left ( \hat{\boldsymbol{y}},\phi_{\ell} ,\theta_{\ell} \right ),\ \forall k\in \mathcal{K}$. Thus, the original problem $\left ( P_\mathcal{A} \right )$ is reduced to only optimizing the transmit beamforming vectors, i.e.,
\begin{subequations}\label{eq24}
\begin{alignat}{3}
&\left ( P_{\mathcal{B}} \right )&\quad &\min_{\left \{ \boldsymbol{w}_{k} \right \}} &\quad& \sum_{k=1}^{K}\left \| \boldsymbol{w}_{k} \right \|^{2} \label{eq24a}\\
& & &\textrm{s.t.} & & {\frac{\left | \boldsymbol{h}_{k}^{H}\left ( \hat{\boldsymbol{y}} \right )\boldsymbol{w}_{k} \right |^{2}}{\sum\limits_{{k}'\neq k}^{K}\left | \boldsymbol{h}_{k}^{H}\left ( \hat{\boldsymbol{y}} \right )\boldsymbol{w}_{{k}'} \right |^{2}+\sigma_{k} ^{2}}\geq \gamma _{k},\quad \forall k\in \mathcal{K}}. \label{eq24b}
\end{alignat}
\end{subequations}

Note that problem $\left ( P_\mathcal{B} \right )$ is the conventional power minimization problem in the multiuser MISO downlink broadcast channel, which can be efficiently solved by utilizing a fixed-point iterative algorithm based on uplink-downlink duality \cite{Book_2001_Bengtsson_Optimum, TSP_2005_Wiesel_Linear}. By leveraging the stationarity of the Karush–Kuhn–Tucker (KKT) conditions \cite{SPM_2014_Bjornson_Optimal}, the optimal minimum mean square error (MMSE) beamforming vectors $\boldsymbol{w}_{k}^{\textrm{o}},\ \forall k\in \mathcal{K}$ have the following general form
\begin{align}\label{eq25}
 \boldsymbol{w}_{k}^{\textrm{o}}=\sqrt{p_{k}}\underbrace{\frac{\left ( \boldsymbol{I}_{N}+\sum\limits_{{k}'=1}^{K}\frac{\lambda _{{k}'}}{\sigma_{{k}'} ^{2}}\boldsymbol{h}_{{k}'}\left ( \hat{\boldsymbol{y}} \right )\boldsymbol{h}_{{k}'}^{H}\left ( \hat{\boldsymbol{y}} \right ) \right )^{-1}\boldsymbol{h}_{k}\left ( \hat{\boldsymbol{y}} \right )}{\left \| \left ( \boldsymbol{I}_{N}+\sum\limits_{{k}'=1}^{K}\frac{\lambda _{{k}'}}{\sigma_{{k}'} ^{2}}\boldsymbol{h}_{{k}'}\left ( \hat{\boldsymbol{y}} \right )\boldsymbol{h}_{{k}'}^{H}\left ( \hat{\boldsymbol{y}} \right ) \right )^{-1}\boldsymbol{h}_{k}\left ( \hat{\boldsymbol{y}} \right ) \right \|}}_{=\tilde{\boldsymbol{w}}_{k}^{\textrm{o}},\textrm{ transmit beamforming direction}},
\end{align}
where $\lambda _{k}\geq 0$ is the Lagrange multiplier corresponding to the $k$-th SINR constraint, while $p_{k}$ denotes the corresponding beamforming power. It can readily be proven that the SINR constraints in \eqref{eq24b} hold with equality at the optimal solution. Since the transmit beamforming directions are already known, solving these $K$ linear equations reveals the values of the $K$ unknown beamforming powers $p_{k}$ as
\begin{align}\label{eq26}
 \left [ p_{1},p_{2},\cdots ,p_{K} \right ]^{T}=\boldsymbol{M}^{-1}\left [\sigma_{1}^{2},\sigma_{2}^{2},\cdots ,\sigma_{K}^{2} \right ]^{T},
\end{align}
where the $\left ( k,{k}' \right )$-th element of the matrix $\boldsymbol{M}\in \mathbb{R}^{K\times K}$ is given by
\begin{align}\label{eq27}
\left [\boldsymbol{M} \right ]_{k,{k}'} = \begin{cases}
\frac{1}{\gamma _{k}}\left | \boldsymbol{h}_{k}^{H}\left ( \hat{\boldsymbol{y}} \right )\tilde{\boldsymbol{w}}_{k} \right |^{2},& k={k}', \\ 
-\left | \boldsymbol{h}_{k}^{H}\left ( \hat{\boldsymbol{y}} \right )\tilde{\boldsymbol{w}}_{{k}'} \right |^{2}, & k\neq {k}'.
\end{cases}
\end{align}

By combining \eqref{eq25} and \eqref{eq26}, we can express the structure of the optimal beamforming vectors in terms of the Lagrange multipliers $\lambda _{1},\lambda _{2},\cdots ,\lambda _{K}$, whose values can be calculated using convex optimization techniques \cite{Book_2001_Bengtsson_Optimum} or by solving fixed-point equations \cite{TSP_2005_Wiesel_Linear}.
\subsubsection{Surface-Shape Morphing ${\boldsymbol{y}}$ Given the Beamforming Vectors $\left \{ {\hat{\boldsymbol{w}}_{k}} \right \}$}
Next, we consider morphing the FIM's surface shape ${\boldsymbol{y}}$, given the calculated transmit beamforming vectors $\hat{\boldsymbol{w}}_{1},\hat{\boldsymbol{w}}_{2},\cdots ,\hat{\boldsymbol{w}}_{K}$ in \eqref{eq25}. In this case, problem $\left ( P_\mathcal{A} \right )$ is reduced to a feasibility-check problem with respect to the FIM's surface shape, which can be formulated as
\begin{subequations}\label{eq29}
\begin{alignat}{3}
&\left ( P_\mathcal{C} \right )&\quad &\textrm{Find} &\quad& \boldsymbol{y} \label{eq29a}\\
& & &\textrm{s.t.} & & \frac{\left | \boldsymbol{h}_{k}^{H}\left ( \boldsymbol{y} \right )\hat{\boldsymbol{w}}_{k} \right |^{2}}{\sum\limits_{{k}'\neq k}^{K}\left | \boldsymbol{h}_{k}^{H}\left ( \boldsymbol{y} \right )\hat{\boldsymbol{w}}_{{k}'} \right |^{2}+\sigma_{k} ^{2}}\geq \gamma _{k},\quad \forall k\in \mathcal{K}, \label{eq29b}\\
& & & & & 0\leq y_{n}\leq y_{\max},\quad \forall n\in \mathcal{N}. \label{eq29c}
\end{alignat}
\end{subequations}

In order to obtain an efficient solution for the FIM's surface shape, we transform problem $\left ( P_\mathcal{C} \right )$ into an optimization problem with an explicit objective function. The underlying rationale is that in the transmit beamforming optimization problem $\left ( P_\mathcal{B} \right )$, all the SINR constraints will be met with equality at the optimal solution \cite{SPM_2014_Bjornson_Optimal}. Therefore, morphing the surface shape to enforce the achievable SINR for each user to be larger than the corresponding SINR target in $\left ( P_\mathcal{C} \right )$ implicitly leads to a lower transmit power value in $\left ( P_\mathcal{B} \right )$. With this in mind, we define a slack variable $\epsilon_k$ to characterize the residual SINR margin for user $k$, i.e.,
\begin{align}\label{eq30}
 \epsilon_k \triangleq \frac{1}{\gamma _{k}\sigma_{k} ^{2}}\left | \boldsymbol{h}_{k}^{H}\left ( \boldsymbol{y} \right )\hat{\boldsymbol{w}}_{k} \right |^{2}- \frac{1}{\sigma_{k} ^{2}}\sum\limits_{{k}'\neq k}^{K}\left | \boldsymbol{h}_{k}^{H}\left ( \boldsymbol{y} \right )\hat{\boldsymbol{w}}_{{k}'} \right |^{2}-1,
\end{align}
which satisfies $\epsilon_k \geq 0,\ \forall k\in \mathcal{K}$. 

By introducing the auxiliary variable $\epsilon_k$, $\left ( P_\mathcal{C} \right )$ is transformed into
\begin{subequations}\label{eq31}
\begin{alignat}{3}
&\left ( P_\mathcal{D} \right )&\quad & \max_{ \boldsymbol{y} } &\quad& \epsilon = \sum_{k=1}^{K}\epsilon _{k} \label{eq31a}\\
& & &\textrm{s.t.} & & 0\leq y_{n}\leq y_{\max},\quad \forall n\in \mathcal{N}, \label{eq31b}\\
& & & & & \epsilon _{k}\geq 0,\quad \forall k\in \mathcal{K}, \label{eq31c}
\end{alignat}
\end{subequations}
which can be efficiently solved by applying the gradient ascent method. Specifically, given the surface shape of the FIM found from the previous iteration as the initial point, we adapt it towards the direction of the gradient for gradually increasing the objective function value $\epsilon$.

According to \eqref{eq30}, the gradient of $\epsilon$ with respect to $\boldsymbol{y}$ is given by
\begin{align}\label{eq32}
 \nabla_{\boldsymbol{y}} \epsilon =&\sum_{k=1}^{K}\frac{1}{\gamma_{k}\sigma _{k}^{2} }\nabla_{\boldsymbol{y}} \left | \boldsymbol{h}_{k}^{H}\left ( \boldsymbol{y} \right )\hat{\boldsymbol{w}}_{k} \right |^{2}\notag \\
 &-\sum_{k=1}^{K}\frac{1}{\sigma _{k}^{2} }\sum_{{k}'\neq k}^{K}\nabla_{\boldsymbol{y}} \left | \boldsymbol{h}_{k}^{H}\left ( \boldsymbol{y} \right )\hat{\boldsymbol{w}}_{{k}'} \right |^{2}.
\end{align}

Note that the gradient $\nabla_{\boldsymbol{y}} \epsilon$ in \eqref{eq32} relies on the general term $\nabla_{\boldsymbol{y}} \left | \boldsymbol{h}_{k}^{H}\left ( \boldsymbol{y} \right )\hat{\boldsymbol{w}}_{{k}'} \right |^{2},\ \forall k, {k}' \in \mathcal{K}$, which can be calculated as \cite{JSAC_2023_An_Stacked}
\begin{align}\label{eq33}
\nabla_{\boldsymbol{y}} \left | \boldsymbol{h}_{k}^{H}\left ( \boldsymbol{y} \right )\hat{\boldsymbol{w}}_{{k}'} \right |^{2} = -2\kappa \sum_{\ell=1}^{L}\sin\theta _{\ell}\sin\phi _{\ell} \notag\\
\times \Im \left \{ \alpha_{k,\ell}\left ( \boldsymbol{a}\left ( \boldsymbol{y},\phi_{\ell} ,\theta_{\ell} \right )\odot \hat{\boldsymbol{w}}_{{k}'}^{\ast } \right ) \boldsymbol{h}_{k}^{H}\left ( \boldsymbol{y} \right )\hat{\boldsymbol{w}}_{{k}'} \right \}.
\end{align}
Substituting \eqref{eq33} into \eqref{eq32} results in an explicit expression of the gradient $\nabla_{\boldsymbol{y}} \epsilon$, which is omitted here for brevity.

Therefore, at each iteration, the surface shape of the FIM is updated by
\begin{align}\label{eq34}
 \boldsymbol{y}\leftarrow \boldsymbol{y} + \mu \nabla_{\boldsymbol{y}} \epsilon,
\end{align}
where $\mu > 0$ represents the step size, which is determined by applying the backtracking line search.

Additionally, a projection process is imposed on each position obtained in \eqref{eq34} to scale it into the effective morphing range of the FIM, yielding
\begin{align}
 y_{n}=\max\left ( \min\left ( y_{n},y_{\max} \right ),0 \right ),\quad \forall n\in \mathcal{N}.
\end{align}

In a nutshell, the proposed alternating optimization algorithm solves problem $\left ( P_\mathcal{A} \right )$ by alternately solving the subproblems $\left ( P_\mathcal{B} \right )$ and $\left ( P_\mathcal{C} \right )$ in an iterative manner. The convergence of the alternating optimization algorithm is guaranteed for two reasons: \emph{i)} the solution for each subproblem ensures that the objective function value of $\left ( P_\mathcal{A} \right )$ is non-increasing over consecutive iterations, and \emph{ii)} the objective function value of $\left ( P_\mathcal{A} \right )$ is lower bounded due to the SINR constraints in \eqref{eq10b}.

\subsection{ZF Beamforming Optimization}
Next, we simplify the transmit beamforming by examining its asymptotically optimal version -- namely zero-forcing (ZF) beamforming -- in the high SNR region. Let us define $\boldsymbol{W} \triangleq \left [ \boldsymbol{w}_{1},\boldsymbol{w}_{2},\cdots, \boldsymbol{w}_{K} \right ]\in \mathbb{C}^{N\times K}$, and $\boldsymbol{H}\left ( \boldsymbol{y} \right ) \triangleq \left [ \boldsymbol{h}_{1}\left ( \boldsymbol{y} \right ),\boldsymbol{h}_{2}\left ( \boldsymbol{y} \right ),\cdots, \boldsymbol{h}_{K}\left ( \boldsymbol{y} \right ) \right ]\in \mathbb{C}^{N\times K}$. The ZF beamformer is given by
\begin{align}\label{eq36}
\boldsymbol{W}_{\textrm{ZF}}=\boldsymbol{H}\left ( \boldsymbol{y} \right )\left ( \boldsymbol{H}^{H}\left ( \boldsymbol{y} \right )\boldsymbol{H}\left ( \boldsymbol{y} \right ) \right )^{-1}\tilde{\boldsymbol{P}}^{1/2},
\end{align}
where $\tilde{\boldsymbol{P}} = \textrm{diag}\left \{ \tilde{p}_{1},\tilde{p}_{2}, \cdots ,\tilde{p}_{K} \right \} \in \mathbb{C}^{K\times K}$ is a diagonal matrix with the $k$-th entry $\tilde{p}_k$ on its diagonal representing the power received at user $k$. To satisfy the constraint in \eqref{eq24b}, we set $\tilde{p}_{k}=\gamma _{k}\sigma _{k}^{2},\ \forall k \in \mathcal{K}$.

Since ZF beamforming projects each user's channel vector $\boldsymbol{h}_{k}\left ( \boldsymbol{y} \right )$ onto a subspace that is orthogonal to the other users' channel vectors, the inter-user interference is completely eliminated, i.e., we have $\boldsymbol{h}_{k}^{H}\left ( \boldsymbol{y} \right )\boldsymbol{w}_{{k}'}=0$ for $\forall {k}'\neq k$. As a result, the objective function $\epsilon_{k}$ in \eqref{eq30} is simplified as
\begin{align}\label{eq37}
 \epsilon_{k,\, \textrm{ZF}} = \frac{1}{\gamma _{k}\sigma_{k} ^{2}}\left | \boldsymbol{h}_{k}^{H}\left ( \boldsymbol{y} \right )\hat{\boldsymbol{w}}_{k} \right |^{2}-1,\quad \forall k \in \mathcal{K}.
\end{align}

Substituting \eqref{eq37} into \eqref{eq31a} and solving problem $\left ( P_\mathcal{D} \right )$, we obtain the surface shape of the FIM with reduced complexity.
\begin{figure}[!t]
\centering
\subfloat[]{\includegraphics[width=4.7cm]{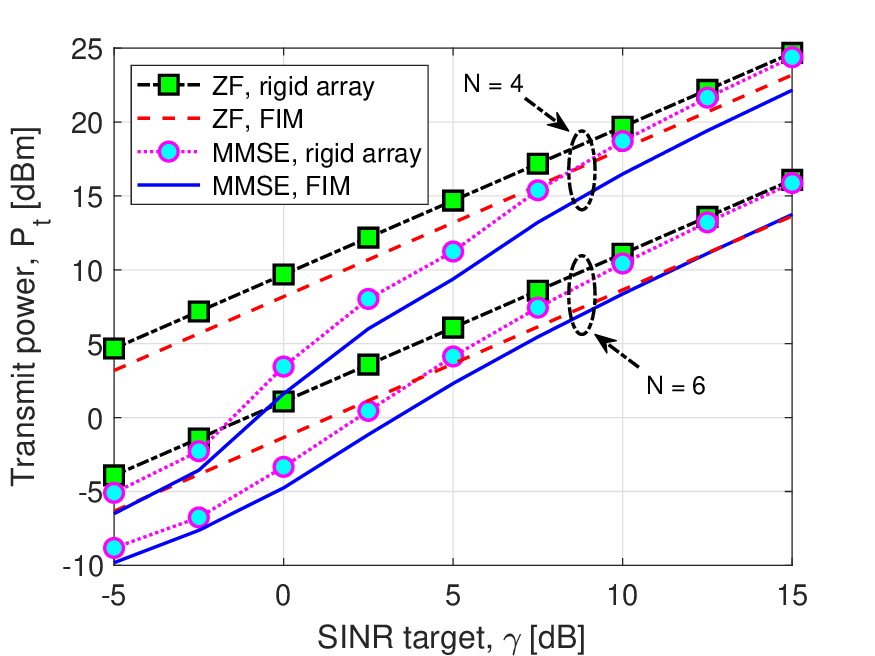}\label{fig_9}}
\subfloat[]{\includegraphics[width=4.7cm]{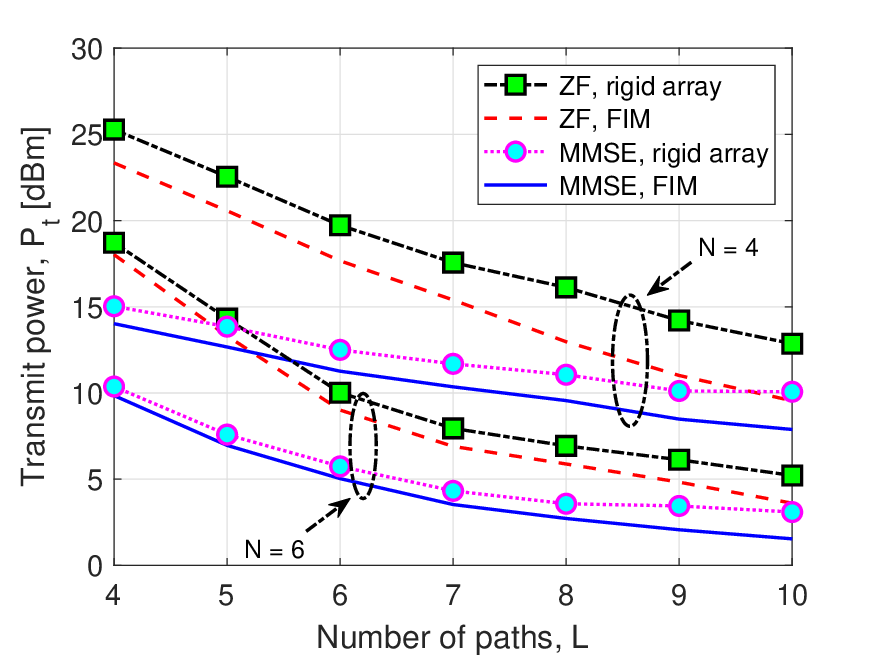}\label{fig_10}}
\caption{(a) Transmit power $P_{\textrm{t}}$ versus the SINR target $\gamma$; (b) Transmit power $P_{\textrm{t}}$ versus the number of propagation paths $L$.}\vspace{-0.1cm}
\end{figure}

\thispagestyle{empty}
\section{Numerical Results}\label{sec6}
In this section, simulation results are presented to verify the benefits of deploying an FIM in wireless networks. An FIM having $N=N_{x}N_{z}$ elements is deployed at the BS with the spacing between adjacent antennas being set to $d_{x}=d_{z}=\lambda /2$. The height of the BS is set to $H_u = 5$ meters (m), and $K$ users are uniformly distributed within a circular region of radius $R_u = 10$ m on the ground. The horizontal distance from the center of the circle to the BS is $D_u = 20$ m. Moreover, $L$ scatterers are uniformly distributed in the far field of the FIM, satisfying $\phi _{\ell}\sim \mathcal{U}\left [ 0,\pi \right )$ and $\theta _{\ell}\sim \mathcal{U}\left [ 0,\pi \right ),\ \forall \ell \in \mathcal{L}$. Additionally, we assume that all the scatterers experience the same channel gain with $\rho _{k,\ell}^{2}=\beta _{k}^{2}/L,\ \forall \ell \in \mathcal{L}$. The distance-dependent path loss is modeled as $\beta ^{2}=\beta_0 ^{2}\left ( d/d_{0} \right )^{-\bar{n}}$ \cite{TCOM_2015_Rappaport_Wideband}, where $\beta_0 ^{2}=\left ( 2\kappa d_{0} \right )^{2}$ is the free space path loss at the reference distance of $d_{0} = 1$ m, and $\bar{n}$ is the path loss exponent, which is set to $\bar{n} = 2.2$ for all users. We assume equal noise power for $K$ users, i.e., $\sigma _{1}^{2}= \sigma _{2}^{2}=\cdots =\sigma _{K}^{2}$. The noise spectral density is $-174$ dBm/Hz \cite{BOOK_2005_Tse_Fundamentals}. The system operates at $28$ GHz with a bandwidth of $100$ MHz. For simplicity, we assume that all the users have identical rate requirements, i.e., $\gamma_{k} =\gamma,\ \forall k \in \mathcal{K}$. Moreover, four transmission schemes are considered to evaluate the performance of the FIM:
\begin{itemize}
 \item ZF/MMSE with a rigid array: The ZF/MMSE transmit beamforming is utilized relying on a rigid 2D antenna array.
 \item ZF/MMSE with an FIM: The surface shape of the FIM and transmit beamforming are alternately optimized. In each iteration, the ZF/MMSE beamforming is designed based on \eqref{eq36}/\eqref{eq25}, while the FIM's surface shape is updated according to \eqref{eq34}.
\end{itemize}
For the proposed alternating optimization algorithm, the maximum tolerable number of iterations is set to $100$, and the convergence is determined if the fractional decrease of the transmit power is less than $-30$ dB. All simulation results are obtained by averaging over $100$ independent channel realizations.

\begin{figure}[!t]
\centering
\subfloat[]{\includegraphics[width=4.7cm]{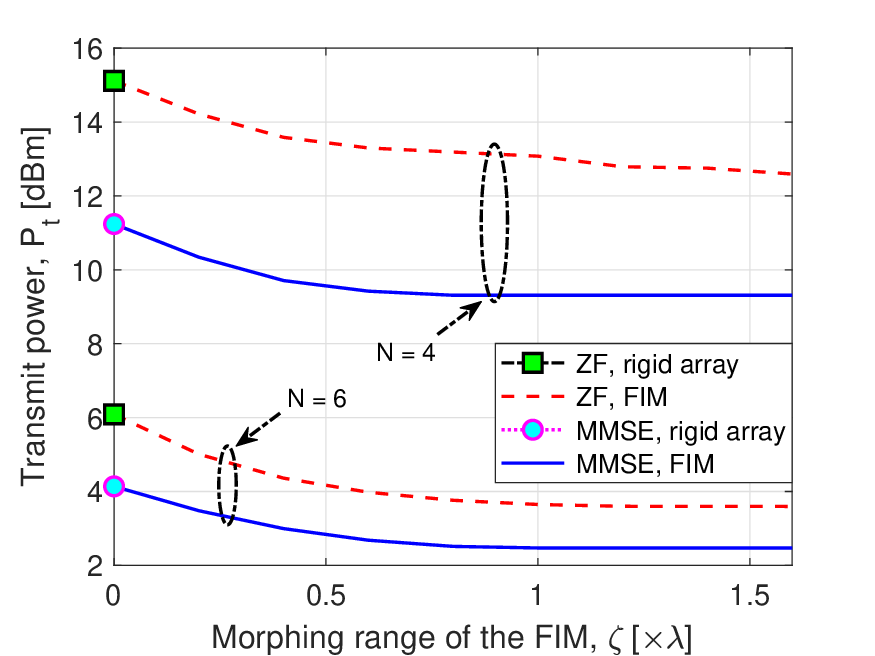}\label{fig_11}}
\subfloat[]{\includegraphics[width=4.7cm]{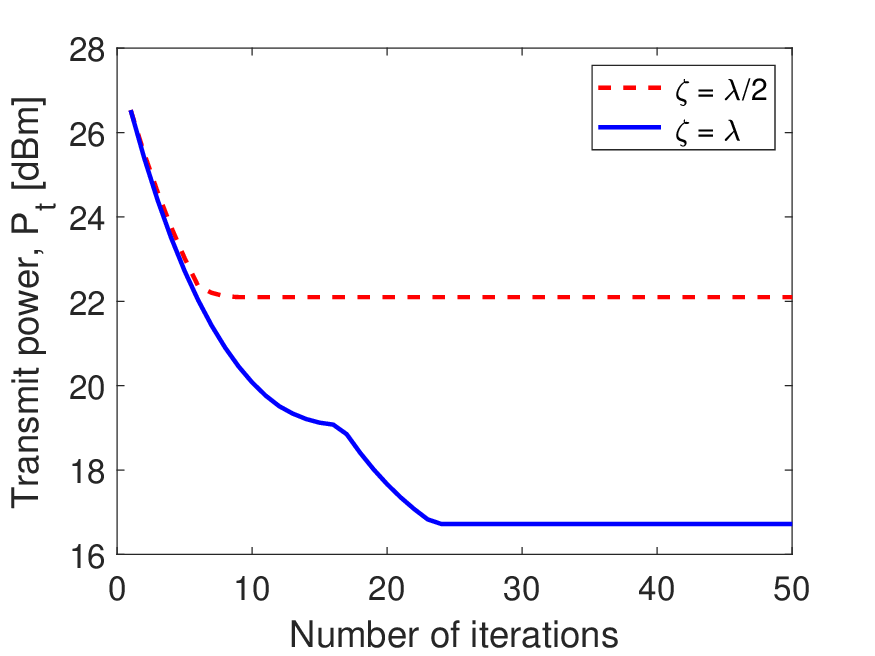}\label{fig_15_1}}
\caption{(a) Transit power $P_{\textrm{t}}$ versus the morphing range of the FIM $\zeta$; (b) Convergence of the transmit power.}\vspace{-0.1cm}
\end{figure}

\thispagestyle{empty}
Fig. \ref{fig_9} shows the transmit power versus $\gamma$ for two different FIM sizes: \emph{i)} $N = 4$, and \emph{ii)} $N = 6$, assuming $N_x = 2$ in both cases. We also assume that there are $K = 4$ users and $L = 8$ propagation paths in the wireless network. The morphing range of the FIM is set to $\zeta = \lambda$. Unsurprisingly, the MMSE beamformer performs best across the entire SNR range, while the ZF beamformer is asymptotically optimal at high SNR, where the noise is dominated by the interference \cite{SPM_2014_Bjornson_Optimal}. As $N$ increases, the transmit power is further reduced, since the $K$ channels become more orthogonal. In all the considered scenarios, the FIM provides an additional SNR gain by morphing its surface shape, further reducing the transmit power by about $3$ dB. Furthermore, Fig. \ref{fig_10} examines the performance as the number of propagation paths in the environment increases. We fix $\gamma = 5$ dB and keep all other parameters the same as in Fig. \ref{fig_9}. It is shown that the performance improves as $L$ increases, since the signal components of multiple propagation paths are more likely to form a favorable profile across the array. Moreover, the 3D FIM outperforms conventional rigid 2D arrays thanks to its morphing capability. Hence, the FIM may beneficially morph to counteract the wireless fading, and its performance gain becomes more significant as the number of propagation paths increases. Fig. \ref{fig_11} plots the transmit power of different schemes as the morphing range $\zeta$ increases, setting $L = 8$ and keeping all other parameters the same. We note that the conventional 2D array is a special case of the FIM when $\zeta = 0$. As $\zeta $ increases, the FIM has more flexibility to adapt its surface shape, gradually reducing the transmit power required. For all setups, an FIM with a morphing range of $\zeta = \lambda$ may reduce the transmit power by about $3$ dB. Moreover, it is interesting to observe that both increasing the number of antennas and the morphing range can narrow the gap between MMSE and ZF beamforming. This is due to the fact that the FIM improves the channel quality for each user and implicitly drives the system to operate in a higher SNR region.

\begin{figure}[!t]
\centering
\subfloat[]{\includegraphics[width=4.7cm]{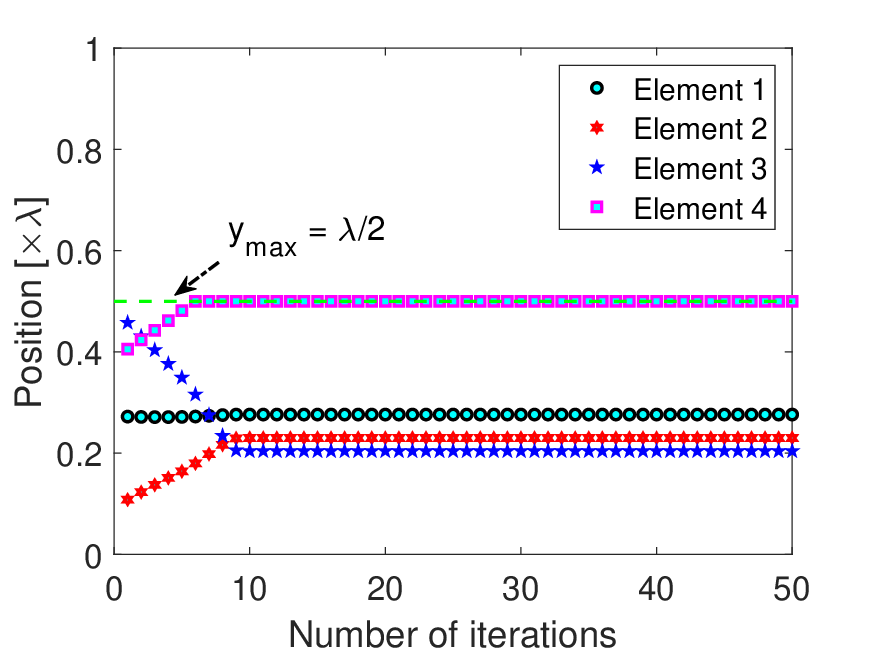}\label{fig_13_1}}
\subfloat[]{\includegraphics[width=4.7cm]{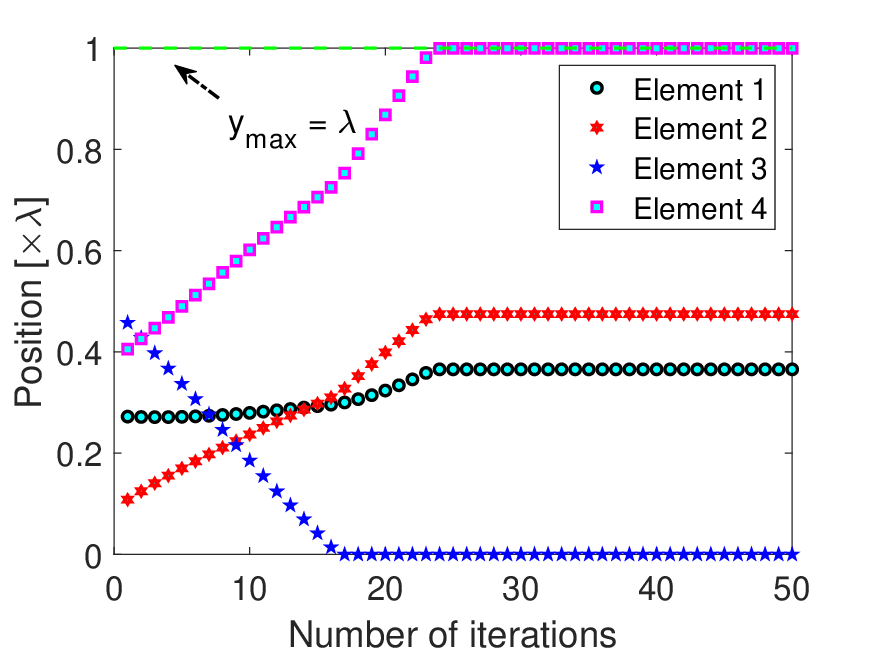}\label{fig_13_2}}
\caption{(a) Convergence of the FIM's surface shape ($L = 4$, and $\zeta = \lambda/2$); (b) Convergence of the FIM's surface shape ($L = 4$, and $\zeta = \lambda$).}\vspace{-0.1cm}
\label{fig_13}
\end{figure}
Finally, we examine the convergence behavior of the proposed alternating optimization algorithm. The simulation setup is the same as Fig. \ref{fig_11}. We consider an FIM having $N = 4$ elements serving $K=4$ users. Specifically, Figs. \ref{fig_15_1} illustrates the evolution of the transmit power, as the iterations proceed, while the corresponding FIM's surface shape is shown in Fig. \ref{fig_13}. Firstly, in Fig. \ref{fig_13_1}, we consider a propagation environment having $L = 4$ scatterer clusters, and the morphing range is $\zeta = \lambda/2$. As expected, the FIM gradually reduces the transmit power, as observed in Fig. \ref{fig_15_1}. However, due to the constrained morphing range, the position of element $4$ increases up to $y_{\max} = \lambda/2$, and the transmit power converges after $10$ iterations. When increasing the morphing range to $\zeta = \lambda$, element $4$ attains an improved morphing range to adapt its position. Interestingly, this also forces elements $1 \sim 3$ to readjust their positions to cooperatively reconfigure the FIM's surface shape, further reducing the transmit power by about $5$ dB. As a result, the transmit power is reduced from $1.4$ dBm to $1.1$ dBm. It is noted that in all setups considered, the proposed alternating optimization method tends to converge within $30$ iterations.

\thispagestyle{empty}
\section{Conclusions}\label{sec7}
FIMs offer unprecedented flexibility in terms of both EM response and mechanical tuning. By morphing its surface shape, we have shown that an FIM deployed at a BS is capable of significantly reducing the power consumption in wireless networks, while maintaining the same QoS requirements. An efficient alternating optimization method has been customized for iteratively optimizing the FIM's surface shape and the transmit beamformer. Numerical results have quantified the performance improvement of using an FIM over conventional rigid 2D arrays. An FIM with a morphing range of one wavelength, which corresponds to $10.8$ mm at $28$ GHz, results in a power gain of at least $3$ dB. However, reaping the performance gain of FIMs in wireless networks also presents significant challenges that must be addressed, such as channel estimation, which constitutes our future research direction.

\section*{Acknowledgments}
This research is supported by the Ministry of Education, Singapore, under its MOE Tier 2 (Award number MOE-T2EP50220-0019).

\bibliographystyle{IEEEtran}
\bibliography{ref}
\end{document}